\documentclass[useAMS,usenatbib]{mn2e}

\usepackage{times}

\usepackage{subfigure}

\usepackage{amsmath}
\usepackage{amssymb}
\usepackage{graphicx}
\usepackage{latexsym}
\usepackage{aas_macros}
\usepackage{subfigure}

\usepackage{url}

\usepackage[squaren]{SIunits}

\newcommand{\mean}[1]{\ensuremath{\langle #1 \rangle}}
\newcommand{\parenfrac}[2]{\paren{\frac{#1}{#2}}}
\newcommand{\paren}[1]{\left ( #1 \right )}
\newcommand{\pderiv}[2]{\frac{\partial {#1}}{\partial {#2}}}
\newcommand{\bracket}[1]{\left [ #1 \right ]}
\newcommand{\beq}{\begin{equation}}
\newcommand{\eeq}{\end{equation}}
\newcommand{\cs}[1]{\ensuremath{c_{\text{s}}^{#1}}}
\newcommand{\led}[1]{\ensuremath{l_{\text{ed}}^{#1}}}
\newcommand{\ved}[1]{\ensuremath{v_{\text{ed}}^{#1}}}
\newcommand{\ted}[1]{\ensuremath{t_{\text{ed}}^{#1}}}
\newcommand{\Ned}[1]{\ensuremath{N_{\text{ed}}^{#1}}}

\newcommand{\fluctuate}[1]{\ensuremath{\paren{#1}_{\text{fl}}}}
\newcommand{\resolution}[1]{\ensuremath{\paren{#1}_{\text{rs}}}}
\newcommand{\Nfl}[1]{\ensuremath{N_{\text{fl}}^{#1}}}
\newcommand{\Nres}[1]{\ensuremath{N_{\text{rs}}^{#1}}}

\newcommand{\Bpeak}{\ensuremath{B_{\nu}^{\text{peak}}}}
\addunit{\AU}{au}

\begin{document}
\title[Imprecision of Accretion Theory  and Protoplanetary Discs]{Quantifying the Imprecision of Accretion Theory  and Implications for
Multi-Epoch Observations of  Protoplanetary Discs}

\author[E.G. Blackman, F. Nauman, R.G. Edgar]
{Eric G. Blackman$^{1}$\thanks{blackman@pas.rochester.edu},
Farrukh Nauman$^{1}$\thanks{fnauman@pas.rochester.edu}, and
Richard G. Edgar$^{1,2}$\thanks{r.g.edgar@gmail.com}, \\
1. Department of Physics and Astronomy, University of Rochester, Rochester, NY 14627\\
2. Massachusetts General Hospital, Martinos Center for Biomedical Imaging Charlestown, MA, 02129}
\date{\today}

\pagerange {\pageref{firstpage}--\pageref{lastpage}}

\label{firstpage}

% ----------------

\maketitle

\begin{abstract}
If accretion disc emission results from turbulent dissipation, then axisymmetric accretion theory must be used as a mean field theory:  turbulent flows are at most axisymmetric only when suitably averaged. Spectral predictions therefore have an intrinsic  imprecision that must be quantified to interpret the variability exhibited by a source observed at different epochs.  We quantify contributions to the stochastic imprecision that come from  azimuthal and radial averaging  and show that the imprecision is minimized for a particular choice of radial averaging, which in turn, corresponds to an optimal spectral resolution of a telescope for a spatially unresolved source.  If  the optimal spectral resolution is less than that of the telescope then the data can be binned to  compare to the theoretical prediction of minimum imprecision. Little stochastic variability is predicted at radii much larger than that at  which the dominant eddy turnover time ($\sim$ orbit time) exceeds the  time interval between observations;  the epochs would then be sampling the same member of the stochastic ensemble.  We discuss the application of these principles to protoplanetary discs for which there is presently a paucity of multi-epoch data but for which such data acquisition projects are underway.
  \end{abstract}

% -------------------


% -------------------
\section{Introduction}
\label{sec:intro}

Accretion discs are ubiquitous in astrophysics, often forming around stars and compact objects where
angular momentum would  inhibit direct gravitational infall of the plasma onto the central object.
A long standing theme of research in accretion theory has been to understand how how the discs transport angular momentum and remain quasi-steady accretors. This transport  likely involves different combination of  local and global mechanisms depending on the circumstances. However, there is presently an intellectual gap between the study of the fundamental theory of angular momentum transport in accretion
discs via numerical simulations and practical accretion disc models
that can be used to compare with observed spectra.
For reviews that sample a range of perspectives
see \cite{2002apa..book.....F,2003LNP...614..329B,2009apsf.book.....H,2010arXiv1005.5279S,2010AN....331..101B}.

%Our understanding of angular momentum transport in discs has not yet progressed to the point where the fruits have produced a practical set of tools for modeling observations beyond that of the the crude approach in \cite{1973A&A....24..337S}, which parameterizes the uncertainty in the mechanism of transport into a single parameter. This approach is commonly used  in practical modeling of observations.

The most commonly used practical accretion disc model  is based on local viscous transport of angular momentum in axisymmetric discs and hides the unknown physics of transport into an effective local ``turbulent viscosity'' \citep{1973A&A....24..337S,1981ARA&A..19..137P,2002apa..book.....F}.
 Because microphysical  molecular viscosity is typically much too small to drive the observed accretion rates given constraints on surface densities, some kind of enhanced transport is needed.
While these mechanisms routinely involve some form of turbulent motion (often with a magnetic origin), the standard practical model equations describing accretion discs do not specify or accommodate
the nuanced physics of transport, lumping the processes into a single dimensionless transport coefficient.
The assumption that the discs are axisymmetric can then at best apply in a spatially or temporally averaged mean-field sense since turbulence necessarily violates axisymmetry locally and turbulent transport is a stochastic process.

In order to properly compare standard axisymmetric accretion theory with observations, the limitations of the  theory must be quantified. The theory has a limited precision which implies that agreement between models and observation can only be approximate. Disagreement within the precision error may be consistent with a stochastic nature of the discs.  A key quantity, and the focus of this paper, is the relative precision error (RPE)\footnote{Compare the statements `babies sleep between one and twenty three hours per day' and `babies sleep \unit{\ensuremath{8\pm1}}{\hour} per day.' The first statement has a large RPE, but is almost certain to be accurate. The second statement has a low RPE, but may well be incorrect. However, its low RPE makes the second statement testable} in the luminosity at a given frequency.
\citet{1998MNRAS.299L..48B} discussed the RPE for accretion discs in the context of Advection Dominated Accretion Flows (ADAFs) and thin discs in active galactic nuclei (AGN) and  X-ray binarys (XRBs) in which the observation times are often longer than the orbit times at a given radius.
In contrast,  protoplanetary discs often involve small observational exposure times compared
to disc orbit times.   In this context, we revisit quantifying the precision of standard accretion disc theory and make predictions for multi-epoch observations. The predictions can help to diagnose whether
 a systematic variability  may also be present.

 In section 2, we compute the azimuthal and radial contributions to the luminosity RPE
  and combine them for a spatially
 unresolved source. In section 3 we discuss how to minimize the RPE and the role of telescope
 spectral resolving power. In section 4 we consider how to apply the results
 to the specific case of LLRL 31, one of the few
 protoplanetary discs in which there are multi-epoch observations.
 Although we find that the observed variability in that source is systematic
 and not stochastic, the application exemplifies how to use the results herein.
 We also discuss further  implications for  future multi-epoch observations.
of protoplanetary discs, and conclude in section 5.

%it is useful to d
%\begin{enumerate}
%\item Unaveraged theoretical accretion discs
%\item Averaged theoretical accretion discs
%\item Observed accretion discs
%\end{enumerate}
%The `averaged theoretical' discs are the well known thin accretion discs discussed by \citet{1981ARA&A..19..137P} (we shall quote results freely from this paper).
%Observed accretion discs are real accretion discs, as observed by instruments such as \emph{Spitzer}.
%These invariably have issues of spatial and spectral resolution associated with them.
%inally, there are the `unaveraged theoretical' discs which are fully self-consistent turbulent disc models.
%Such a model would be the closest approximation to a `real' disc we could construct.

%

\section{Calculation of  RPE  in Luminosity from Turbulent Dissipation }

We assume a standard $\alpha$-disc prescription \citep{1973A&A....24..337S} for the turbulent viscosity.The characteristic turbulent eddy scales have characteristic velocities \ved{} and sizes \led{} so that the effective viscosity satisfies
\begin{equation}
\nu \approx \alpha \cs{} h \approx \ved{} \led{} \sim {v_{ed}^2\over \Omega}
\label{1}
\end{equation}
where \cs{} is the disc sound speed, $h$ is the scale height,
and $\Omega$ is the orbital speed.
We take $0<\alpha<1$ to be a constant over the whole disc.
The last similarity in (\ref{1}) follows from the assumption that the
eddy turnover time $\tau_{ed}\sim {1\over \Omega}$  consistent with simulations of the magnetorotational instability
(MRI) (e.g. \cite{2003LNP...614..329B}). For a thin disc in hydrostatic equilibrium, Eq. (\ref{1}) also implies that
$v_{ed} \sim \alpha^{1/2} c_s$ and $l_{ed} \sim \alpha^{1/2} h$ \citep{1998MNRAS.299L..48B}.

In what follows, we focus on cases for which the
required observational exposure time is much less than the eddy
turnover time at all radii. We can then regard a measured spectrum as
 an instantaneous snapshot.
This applies  to \emph{Spitzer} observations of protoplanetary discs.
The discs range in size from a few tenths of an AU{} to several tens of AU
so that \ted{} ranges from days (in the inner disc) to years (in the outer disc).
In contrast, a typical \emph{Spitzer} exposure takes only a few minutes.
%The role of observation times longer than eddy turnover times woudl add an extra
%factor.

Since the sources of interest are spatially unresolved,
we calculate the contributions to the RPE by
averaging in the azimuthal and in the radial directions for the axisymmetric theory.
If  the disc is sufficiently optically thick then  the photosphere
occurs within the first layer of energy containing eddies.
An optically thin disc requires additionally averaging over the third dimension of eddies.

\subsection{Azimuthal Contribution}

\label{sec:simple}

Consider an annulus of width equal to one eddy scale centered at a radius $r$ within the disc.
The number of eddies producing observable emission in this  annulus is given by
\begin{equation}
\Ned{} \sim  \frac{2 \pi r}{\led{}}  \left[ 1+\left({h\over l_{ed} }-1\right)e^{-\tau_{\nu} }\right],
%\label{eq:NedDefine}
\label{2}
\end{equation}
where $\tau_\nu$ is the optical depth. In the limit that $\tau_\nu \rightarrow 0$, eddies over the entire thickness of the disc contribute to the emission. For $\tau_\nu \rightarrow \infty$ only the top layer of eddies contribute
to the emission.

Over its lifetime, the brightness of each dominant  eddy will rise to some maximum and then
fall as it cascades to small scales. Let us assume that the rate of brightening and dimming are constant so
that the eddy brightens to its peak luminosity $L_0$ at  $t_{ed}/2$ and then falls to zero
at $t=t_0$.
The probably distribution function for the luminosity during the rise and fall is then
\begin{equation}
p(L_e) d{L_e} = \frac{1}{L_0}dL_e.
\end{equation}
Then the mean luminosity for each eddy is $\langle L_e \rangle =L_0/2$
and the
mean squared luminosity is
\beq
 {\langle L_e^2\rangle} = {1\over 2}\left[
  {\int_0^{L_0}  {L_e^2\over L_0} dL_e
\over \int_0^{L_0} {1\over L_0} dL_e } +
 {\int_{L_0}^0  {L_e^2\over L_0} dL_e
\over \int_{L_0}^0 {1\over L_0} dL_e }\right]
= L_0^2/3.\eeq
The variance is then
\begin{equation}
\sigma^2(L_e) = \mean{L_e^2} - \mean{L_e}^2 = \frac{L_0^2}{12}
\end{equation}

If the value of the mean luminosity of each eddy is normally distributed about
$L_0/2$ then the standard deviation of the luminosity per eddy when averaged over
all $N_{ed}$ eddies in a given annulus satisfies $ \frac{\sigma(L)}{\sqrt{\Ned{}}}$.
Using $L_\nu$ to indicate the luminosity at frequency $\nu$ measured from the
annulus whose  peak temperature corresponds to that  frequency, the fractional luminosity variation of the entire annulus of $N_{ed}$ eddies is then
\begin{equation}
\left[\frac{\Delta L_\nu}{{L_\nu}}\right]_{az}
=\frac{N_{ed}\Delta L_e}{N_{ed}\mean{L_e}}
= {\sigma (L_e) / {\sqrt N_{ed}} \over   L_0/2}
= \frac{1}{\sqrt{3 \Ned{}}}
%,
\label{5}
\eeq
where we have used Eq. (\ref{2}),  the relations below Eq. (\ref{1}), and the subscript
$az$ to indicate the azimuthal contribution to the variation.
In the optically thick limit Eq. (\ref{5}) gives
$\left[\frac{\Delta L_\nu}{{L_\nu}}\right]_{az}=\frac{\alpha^{\frac{1}{4}}}{\sqrt{6 \pi}} \parenfrac{h}{r}^{\frac{1}{2}}$.

%
%
%
%In Figure~\ref{fig:SimpleObs}, we show a schematic of the expected effect of the turbulence on observations.
%The thick solid line is the theoretical spectrum, as predicted by standard thin disc theory.
%Section 3.4 of \citet{1981ARA&A..19..137P} shows that the slope of this line would be $L_{\nu} \propto \nu^3$ (and we need a new symbol here, to avoid confusion with viscosity).
%The dotted lines denote the 2\% variability limits we have just discussed.
%Finally, the curves are two possible observations of the accretion disc, separated by a year or so.
%At the low frequency end, which probes the outermost region of the disc, the lines are identical.
%As noted above, the eddy turnover timescale is roughly the orbital timescale, so eddies at \unit{10}{\AU} or so will be effectively unchanged between observations.
%At the high frequency end, the curves are dramatically different, while still lying within the 2\% envelope.
%The high frequency end probes the inner portions of the disc, where the orbital timescales (and hence \ted{}) are days.
%After a year, \emph{Spitzer} will be observing a completely new set of eddies, and hence we should not expect the curves to be similar.
%We can only expect the spectrum to be confined by the 2\% envelope.
%
%
%
%
%
%-------------------------------------------------------------------------------------------------------
%

\subsection{Radial Contribution}
\label{sec:radial}
%
%

%The eddies in an accretion disc azimu
%In reality, they will have a radial component - roughly equal to $h \sqrt{\alpha}$.
%This is going to affect the emitted luminosity, and we will now discuss a method of calculating this.
%
%
%
In determining the radial contribution to the luminosity RPE we
consider two independent  contributions. The first comes from
a radial smoothing implicit to the mean field theory. The second comes
from turbulent velocity fluctuations that add macroscopic fluctuations about the mean orbital
velocity at each radius. We address each in turn.

In formulating a mean field theory for accretion discs,
one can in  principle simply take vertical and azimuthal averages to obtain a theory that depends only
on radius without any radial smoothing.  However in the  commonly employed
 framework which invokes a turbulent viscosity
 (Shakura \& Sunyaev 1973),  the scale of the turbulence is assumed to be
smaller than the scale of the mean quantities.  The theory is not really resolved on spatial
scales below an eddy scale so we therefore include a
radial smoothing scale, $\xi$, chosen such that $\led{} < \xi < r_{\text{disc}}$.
In addition, telescopes have a finite spectral resolution and a given frequency  in practice corresponds to a radius range in the disc, not a precise radius for a spatially unresolved source.
We therefore invoke a radial smoothing  of mean quantities of the form
\begin{equation}
{\overline X}(r)
%\mean{X(r)}_{\text{space}}
       \approx \mean{X(r,t)}_{\xi}
       = \int_{-\xi}^{\xi} X(r+\lambda,t) d \lambda,
\label{6}
\end{equation}
where $X$ is an arbitrary quantity to be averaged and $\overline X$ is its mean.
The similarity follows  from the assumption that eddy statistics are steady in time.
Blackman (1998) discussed an additional temporal average but
as mentioned above, our present interest is for cases in which the required snapshot observation time is much less than an eddy turnover time at all radii.
%This applies  to  \emph{Spitzer} observations of protoplanetary discs.

%
%
%
%
The radial contribution to the luminosity variation can be written as
\begin{equation}
\left[\frac{\Delta L_{\nu}}{L_{\nu}}\right]_{rad}
\approx
\left | \frac{r}{L_{\nu}} \pderiv{L_{\nu}}{r} \right|
\frac{\Delta r}{r}.
\label{eq:VariabilityEquation}
\end{equation}
We must now calculate the contributions to the right hand side.

%\subsubsection{Calculation of $\Delta r/r$}
We write the total imprecision in radius  as a quadratic sum of
 the smoothing and velocity fluctuation contributions. That is,
\begin{equation}
\frac{(\Delta r)^{2}}{r^{2}} \approx
\frac{\resolution{\Delta r}^{2}}{r^{2}}+
\frac{\fluctuate{\Delta r}^{2}}{r^{2}}
\label{eq:radiusRPEdefine}
\end{equation}
where first term comes from the blurring induced by the radial smoothing scale $\xi$  and the second term comes from the macroscopic turbulent velocity fluctuations.
%\texttt{Why these don't add in quadrature, I'm not sure\ldots}

For the radial smoothing term of  Eq.  (\ref{eq:radiusRPEdefine}),
we use the fact that
\begin{equation}
r - \frac{\xi}{2 \Nres{\frac{1}{2}}} < r <
r +\frac{\xi}{2 \Nres{\frac{1}{2}}}
\end{equation}
are indistinguishable once $\xi$ is chosen, so that
\beq
{(\Delta r)_{rs}\over r}= \frac{\xi}{2r \Nres{\frac{1}{2}}},
\label{12}
\eeq
where
\begin{equation}
\Nres{} = 1 +\left({h\over l_{ed} }-1\right)e^{-\tau_{\nu}}
+ \frac{t_{\text{obs}}}{\ted{}}
%\sim  1
%+\left({h\over l_{ed} }-1\right)e^{-\tau_{\nu}}
\label{12a}
\end{equation}
and measures the effective number of averaging scales per $\xi$. The role of $\tau_\nu$  the same as that for Eq.
(\ref{2}). In the limit $t_{obs}<<t_{ed}$ and $\tau_\nu >> 1$, $N_{fl} =1$.
For the velocity fluctuation contribution to Eq. (\ref{eq:radiusRPEdefine}),
we note that  the dominant contribution to the local velocity at a given radius is  $\propto r^{-1/2}$ from
Keplerian rotation. Thus
\begin{equation}
\frac{\fluctuate{\Delta r}}{r} \approx 2 \frac{\fluctuate{\Delta v_0}}{v_0} \approx 2 \frac{\ved{}}{v_0 \Nfl{\frac{1}{2}}},
\label{9}
\end{equation}
where we have again used the standard result for error on the mean as applied to $N_{fl}$ eddies each of whose mean velocity is normally distributed with fluctuations of order $v_{ed}$.
  The number of eddies over which the radial averaging is performed  is given by
\begin{equation}
\Nfl{} = \frac{\xi}{\led{}}\left[1+\left({h\over l_{ed} }-1\right)e^{-\tau_{\nu}} \right]
+ \frac{t_{\text{obs}}}{\ted{}}
%\sim\frac{\xi}{\led{}}
%\sim\frac{\xi}{\led{}}\left[1+\left({h\over l_{ed} }-1\right)e^{-\tau_{\nu}} \right],
\label{eq:NflDefine}
\end{equation}
For $\tau_\nu >> 1$ and
%because although
 %a long exposure would increase $N_{fl}$
%from the snapshot value $\xi / \led{}$,
%For this reason, we have included the second term in Equation~\ref{eq:NflDefine}.
%However, as noted above
%for IR observations of circumstellar discs
 %the observations are effectively snapshots
 %so
  $t_{\text{obs}}/t_{ed}\sim 0$, this gives $N_{fl}= \xi/l_{ed}$.
In this limit, adding (\ref{12}) and (\ref{9}) in quadrature, we obtain
\begin{equation}
\frac{(\Delta r)^{2}}{r^{2}} \approx
4 \frac{v_{\text{ed}}^{2} l_{\text{ed}}}{v_{0}^{2} \xi}
+ \frac{\xi^{2}}{4 r^{2}}.
\label{eq:deltarexpand}
\end{equation}

%\subsubsection{Calculation of $\partial L_{\nu}/\partial r$}
If each ring of material radiates around a specific blackbody temperature $T$,  the luminosity of an annulus of
width  $l_{ed}$ is
\begin{equation}
L_{\nu}  = 2 \pi r \Bpeak{}\paren{T(r)} l_{ed},
\label{eq:LnuDefine}
\end{equation}
where
\beq
B_{\nu}^{\text{peak}}(T)=A T^3
\label{16a}
\eeq
 is the peak value of the Planck function $B_\nu$ and
$A$ is a constant that comes from evaluating
%\frac{\xi}{2 \Nres{\frac{1}{2}}}
\begin{equation}
B_{\nu}
(T) = T^3 \frac{2 k^3}{h^2 c^2} \frac{x^3}{e^x -1}
\end{equation}
at  $x \equiv h \nu / k T
\sim 2.82$,  corresponding to the Wien displacement law.
Then
\begin{equation}
\pderiv{L_{\nu}}{r}
% 2 \pi l_{ed} \bracket{\Bpeak{} + r \pderiv{T}{r} \pderiv{\Bpeak{}}{T}}
=  2 \pi A T^2 \bracket{l_{ed}T + 3 r l_{ed}\pderiv{T}{r}+ r T {dl_{ed}\over dr}}.
\label{eq:pderivLnu}
\end{equation}
If we assume that
 \begin{equation}
T = C r^{-p},
\label{eq:TradialPowerLaw}
\end{equation}
then  combining with the scalings below Eq. (1) gives  $l_{ed}=\alpha^{1/2}h\propto c_s/\Omega \propto r^{{1\over 2}(3-p)}$.
%(approximate for a thin disc at least over a scale $\Delta r < r$)
In combination  with Eqs~(\ref{eq:VariabilityEquation}), (\ref{eq:LnuDefine}) and (\ref{eq:pderivLnu}), we then obtain
\begin{equation}
\left[\frac{\Delta L_{\nu}}{L_{\nu}}\right]_{rad}
\approx
\left| 1 - 3p +{3\over 2}-{p\over 2} \right|
\frac{\Delta r}{r}=\left| {5\over 2}-{7p\over 2} \right|
\frac{\Delta r}{r}
\label{eq:DiscVariability}
\end{equation}
\citet{2007MNRAS.381.1280E} calculated $p$ for a variety of disc models, and  their values are given in Table~\ref{tbl:TemperaturePowerLaws}.
Eq. (\ref{eq:DiscVariability}) is strictly applicable  for the $\tau_\nu >>1$ regime so
here $p=9/10$.
%EGB our predictions are only for turbulence produced by stochastic process. have to think about p from the reflected process.

%Note that the constant $C$ has dropped out of this equation.
%Only the radial power, $p$, is relevant.
%It is somewhat perturbing to note that $p=1/3$, where the temperature is set by irradiation from the central star (likely to be the case for much of the outer disc), has an implied variation of zero.
%
%
\begin{table}
\begin{tabular}{r|l}
Disc Model  & $p$ \\
\hline
Viscously Heated, Optically Thick  & $9/10$  \\
Viscously Heated, Optically Thin   & $3/5$ \\
Irradiated by central star   &  $1/3$
\end{tabular}
\caption{Values for $p$ in Eq. (\ref{eq:TradialPowerLaw}) for various disc models \citep{2007MNRAS.381.1280E}}.
\label{tbl:TemperaturePowerLaws}
\end{table}
\section{Total RPE and Role of  Resolving Power}

Eqs. (\ref{5}) and (\ref{eq:DiscVariability})
give the respective  azimuthal and radial contributions to the
RPE of the predicted luminosity in terms of disc model properties. They must be added in quadrature
to get the total luminosity RPE. The choice of $\xi$ can be taken to minimize the RPE but whether this minimum
can be compared with data depends on the resolving power of the instruments.
We explain these points further below.

\subsection{Minimizing the Intrinsic RPE}
%Imprecision}
First we determine the minimum  of Eq. (\ref{eq:deltarexpand}),
which applies in the optically thick $\tau_\nu >>1$ and $t_{obs}=0$ limit.
Differentiating and setting equal to zero gives
\begin{equation}
\frac{\xi_{\text{opt}}}{r} = 2 \parenfrac{\ved{}}{v_0}^{\frac{2}{3}} \parenfrac{\led{}}{r}^{\frac{1}{3}}.
\label{eq:xioptdefine}
\end{equation}
%\texttt{There's an extra 2 in this equation as compared with Eric's paper}
The optically thin case gives the same result because the optically thin corrections to $N_{rs}$ and $N_{fl}$
just produce an overall multiplicative factor of $l_{ed}/h$ to both terms in
Eq. (\ref{eq:deltarexpand}) and thus do not change its minimum.

Choosing $\xi=\xi_{opt}$ will minimize the RPE
of the theory.  Increasing $\xi$ would average over more turbulent fluctuations, but at the price of poorer spatial precision. Reducing $\xi$ would increase the random error due to a smaller number of eddies.
If  $\xi_{\text{opt}} < \led{}$,  we must use $\xi = \led{}$ instead.

Substituting Eq. (\ref{eq:xioptdefine}) into Eq. (\ref{eq:deltarexpand}), gives
\begin{equation}
\left . \frac{\Delta r}{r} \right |_{\text{min}}
=
\sqrt{3}
\parenfrac{\ved{}}{v_0}^{\frac{2}{3}}
\parenfrac{\led{}}{r}^{\frac{1}{3}},
%=\sqrt{3}\parenfrac{2 \ved{}}{v_0}^{\frac{1}{3}}\parenfrac{2 \alpha \cs{2}}{\Omega^2 r^2}^{\frac{1}{3}}
\label{eq:deltarOpt}
\end{equation}
where we have used  $\ved{} \led{} = \alpha \cs{} h$.
Note that  $\xi_{\text{opt}}$ of Eq. (\ref{eq:xioptdefine})
can only be used to compare with observations if the observations are resolved on
scales $\xi\le \xi_{opt}$ since only then can the data be binned accordingly.
We will come back to this point. First we calculate the RPE for the luminosity using
$\xi =\xi_{opt}$.

%We expect that $\led{}$ and $\ved{}$ will be controlled by $\alpha$, which measures the turbulence within the disc.
%In particular, most work takes
%\texttt{Reference?}
%\begin{eqnarray}
%\led{} & = & \alpha^{q} h \\
%\ved{} & = & \alpha^{1-q} \cs{}
%end{eqnarray}
%%for some $0 < q < 1$.
Substituting $l_{ed}= \alpha^{1/2}h$ and $v_{ed}=\alpha^{1/2}c_s$ into
Eq. (\ref{eq:deltarOpt}), and in turn substituting the result into Eq. (\ref{eq:DiscVariability})
gives
\begin{equation}
%\left . \frac{\Delta L_{\nu}}{L_{\nu}} \right |_{\text{rad,opt}} \approx \sqrt{3} \left| {5\over 2}-{7p\over 2} \right|\alpha^{\frac{2-q}{3}}
\left . \frac{\Delta L_{\nu}}{L_{\nu}} \right |_{\text{rad,opt}} \approx \sqrt{3} \left| {5\over 2}-{7p\over 2} \right|\alpha^{\frac{1}{2}}
\parenfrac{h}{r}.
\label{eq:LvaryOpt}
\end{equation}
%We shall calculate in more detail below, but we can expect $p \approx 1$, $q=0.5$, $\alpha=10^{-3}$ and $h/r \approx 0.1$ for a typical protoplanetary disc.
%In this case, the fractional variability turns out to be around 1\%.
%This is similar to the value calculated in Section~\ref{sec:simple}, which is reassuring.
%Note that if $\alpha \approx 0.1$, which is often claimed, then Equation~\ref{eq:LvaryOpt} will become closer to 10\%, which is a very significant change.
%
Combining results from Eqs. (\ref{5}) and (\ref{eq:LvaryOpt}) by adding the radial and azimuthal
contributions in quadrature,we have for the net minimum RPE in the $\tau_\nu >>1$ limit
\begin{equation}
%\left . \frac{\Delta L_{\nu}}{L_{\nu}} \right |_{\text{opt}}\approx\left[{{3}\over 4}( 5-7p)^2\alpha^{\frac{4-2q}{3}}\parenfrac{h}{r}^2 + {\alpha^{q}\over 6\pi} \parenfrac{h}{r}\right]^{1/2}
\left . \frac{\Delta L_{\nu}}{L_{\nu}} \right |_{\text{opt}}\approx\left[{{3}\over 4}( 5-7p)^2\alpha \parenfrac{h}{r}^2 + {\alpha^{1/2}\over 6\pi} \parenfrac{h}{r}\right]^{1/2},
\label{eq:LvaryOptCom}
\end{equation}
where in the optically thick limit \citep{2007MNRAS.381.1280E} $p=9/10$ and
 $h/r$ is
 given by
 \begin{equation}
\frac{h}{r}  =
1.8 \times 10^{-2} \left(\frac{\dot{M}_{\ast}}{10^{-8} M_{\odot} yr^{-1}} \right)^{1\over 5} \left(\frac{M_{\ast}}{M_{\odot}} \right)^{-7\over 20} \left(\frac{r}{AU} \right)^{1\over 20}.
\label{depth}
\end{equation}

%As an example, for $p \approx 1$,
%%$q=0.5$,
%and $\alpha=10^{-3}$ and $h/r \approx 0.1$
%for a typical protoplanetary disc, we obtain  $\left . \frac{\Delta L_{\nu}}{L_{\nu}} \right |_{\text{opt}}\sim  1.4\%$. For   $\alpha = 0.1$, we obtain instead  6.8\%.
%The RPE  is depends on
%To review what we have calculated here.
%We are assuming that the `real' disc is turbulent, and that we are attempting to construct an $\alpha$-disc model which best describes the turbulent disc.
%To do this, we average over a particular range of disc annuli, and in this section we assume that we are free to choose the range of annuli averaged ($\xi$).
%In this case, Equation~\ref{eq:LvaryOptCom} tells us the stochastic variability we should expect if we construct the `best' possible $\alpha$-disc model.
%
%
%
% ------------------------------------------------------------
%
\subsection{Expressing Total RPE as function of Resolving Power}

For a telescope, we can only choose $\xi$ to be as small as the  spatial or spectral
resolution allows. If  the smallest
$\xi$ that the telescope can resolve is smaller than $\xi_{opt}$, the data can be binned
to create an effective $\xi=\xi_{opt}$ that minimizes the RPE.

We can estimate the radial averaging scale from the spectral resolving power of the telescope.
The resolving power is given by
\begin{equation}
R \equiv \frac{\nu}{\Delta \nu}.
\end{equation}
From the Wien displacement law, we know that a blackbody spectrum has $\nu_{\text{peak}} \propto T_{\text{peak}}$, so the effective resolving power is then  $R_{ef} \approx T / \Delta T$.
From Eq. (\ref{eq:TradialPowerLaw}), we then obtain
\begin{equation}
|\xi_{\text{ef}}| \approx \frac{r}{pR_{\text{ef}}}
\label{eq:xirealdefine}
\end{equation}
Substituting into Eq. (\ref{eq:deltarexpand}),  we find
\begin{equation}
%\left . \frac{\Delta r}{r} \right |_{\text{ef}}\approx\left( 4 \alpha^{2 - q} \parenfrac{h}{r}^{3} pR_{\text{ef}} + \frac{1}{4p^{2}R^{2}_{\text{ef}}} \right)^{\frac{1}{2}}
\left . \frac{\Delta r}{r} \right |_{\text{ef}}\approx\left( 4 \alpha^{3/2} \parenfrac{h}{r}^{3} pR_{\text{ef}} + \frac{1}{4p^{2}R^{2}_{\text{ef}}} \right)^{\frac{1}{2}}
\label{eq:deltarReal}
\end{equation}
Using this in (\ref{eq:DiscVariability}) we obtain
\begin{equation}
%\left . \frac{\Delta L_{\nu}}{L_{\nu}} \right |_{\text{rad,ef}}\approx \left |{ 5\over 2}-{7p\over 2 }\right |\left( 4 \alpha^{2 - q} \parenfrac{h}{r}^{3} pR_{\text{ef}} + \frac{1}{4p^{2}R^{2}_{\text{ef}}} \right)^{\frac{1}{2}}.
\left . \frac{\Delta L_{\nu}}{L_{\nu}} \right |_{\text{rad,ef}}\approx \left |{ 5\over 2}-{7p\over 2 }\right |\left( 4 \alpha^{3/2} \parenfrac{h}{r}^{3} pR_{\text{ef}} + \frac{1}{4p^{2}R^{2}_{\text{ef}}} \right)^{\frac{1}{2}}.
\label{eq:LvaryReal}
\end{equation}
The azimuthal contribution Eq. (\ref{5}), does not depend on the frequency explicitly
so combining it in quadrature with Eq. (\ref{eq:LvaryReal})
gives  the total effective RPE
\begin{equation}
\begin{array}{r}
%\left . \frac{\Delta L_{\nu}}{L_{\nu}} \right |_{\text{ef}}\approx\left[\left |{5\over 2}-{7p\over 2 }\right |^2\left(4 \alpha^{2 - q} \parenfrac{h}{r}^{3} pR_{\text{ef}} + \frac{1}{4p^{2} R^{2}_{ef}} \right) + {\alpha^{q} \over 6 \pi}\left({h \over r}\right) \right]^{1\over 2}
\left . \frac{\Delta L_{\nu}}{L_{\nu}} \right |_{\text{ef}}\approx\left[\left |{5\over 2}-{7p\over 2 }\right |^2\left(4 \alpha^{3/2} \parenfrac{h}{r}^{3} pR_{\text{ef}} + \frac{1}{4p^{2} R^{2}_{ef}} \right) + {\alpha^{1/2} \over 6 \pi}\left({h \over r}\right) \right]^{1\over 2}.
\label{eq:LvaryRealCom}
\end{array}
\end{equation}
%Evaluating this with $R_{ef}=200$
%EGB Check if this should  100 or perhaps use 200
%and reasonable values for a protoplanetary disc again with $\alpha = 10^{-3}$ gives an error of 7.6\%, while $\alpha = 0.1$ gives an error of 40\%.
It might seem surprising that increasing $R$ in Eq. (\ref{eq:LvaryReal})
 can give less precise results but recall that for a fixed $p$ and $r$ there is a one-to-one correspondence between $R_{ef}$ and $\xi$.  But as discussed above, there is a value $\xi=\xi_{opt}$  that minimizes the RPE.
There is therefore a corresponding value of $R_{ef}$  that minimizes the RPE and the two turn out to be
related by  $\xi_{opt} =  {r\over p R_{opt}}$.
%Averaging over too small a scale or using too high a resolving power has less error reduction than averaging of a  larger scale, which samples more eddies.  But averaging over too large a scale then blurs together values from different scales.  Thus there is an optimal value of $\xi$ or $R_{ef}$ that minimizes the RPE.
Figure 1 illustrates these basic principles in  plots of  Eq. (\ref{eq:LvaryRealCom})
for different values of $\alpha$. Note that the minima are less pronounced as $\alpha$ is decreased.

Note that  if $R$ of the telescope is sufficiently large and the data are not more coarsely binned
by hand, the data could represent a radial resolution less than the eddy scale of the theory. The  $\alpha$-disc model based on turbulence becomes ill-defined below such scales.
Coarse graining of the data has to be considered carefully  before sensibly comparing
with a mean field theory.

\subsection{Optically thin case}
Finally, note that
for $\tau_\nu<<1 $  Eqs. (\ref{2})
(\ref{12a}) and (\ref{eq:NflDefine}) take a different values compared to the $\tau_\nu >>1$ case.
The overall calculation of the RPE follows that which led to (\ref{eq:LvaryRealCom}) with the analogous
result being
\begin{equation}
\begin{array}{r}
\left . \frac{\Delta L_{\nu}}{L_{\nu}} \right |_{\text{ef}}\approx\left[\left |{5\over 2}-{7p\over 2 }\right |^2\left(4 \alpha^{2} \parenfrac{h}{r}^{3} pR_{\text{ef}} + \frac{\alpha^{1/2}}{4p^{2} R^{2}_{ef}} \right) + {\alpha \over 6 \pi}\left({h \over r}\right) \right]^{1\over 2}.
\label{30thin}
\end{array}
\end{equation}
where in this $\tau_\nu <<1$ limit \citep{2007MNRAS.381.1280E}, $p=3/5$ and
\begin{equation}
\frac{h}{r}  =
2.8 \times 10^{-2} \left(\frac{\dot{M}_{\ast}}{10^{-8} M_{\odot} yr^{-1}} \right)^{1\over 10} \left(\frac{M_{\ast}}{M_{\odot}} \right)^{-2\over 5} \left(\frac{r}{AU} \right)^{1\over 5}.
\label{depththin}
\end{equation}
%Evaluating this with $R_{ef}=200$ and reasonable values for a protoplanetary disc again with $\alpha = 10^{-3}$ gives an error of 1.4\%, while $\alpha = 10^{-1}$ gives an error of 16\%.
To obtain minimum (optimal) value of (\ref{30thin}) (the analogue of (\ref{eq:LvaryOptCom})),
we substiute in (\ref{30thin}) the value
$R_{ef}= R_{opt}= {r/p \xi_{opt}}$
and use $\xi_{opt}/r $ from Eq. (\ref{eq:xioptdefine})
which applies for both the optically thin and thick cases,

%= 2\parenfrac{\ved{}}{v_0}^{\frac{2}{3}} \parenfrac{\led{}}{r}^{\frac{1}{3}}$, whichis the optically thin version of  of

\begin{figure}
\centering
\includegraphics{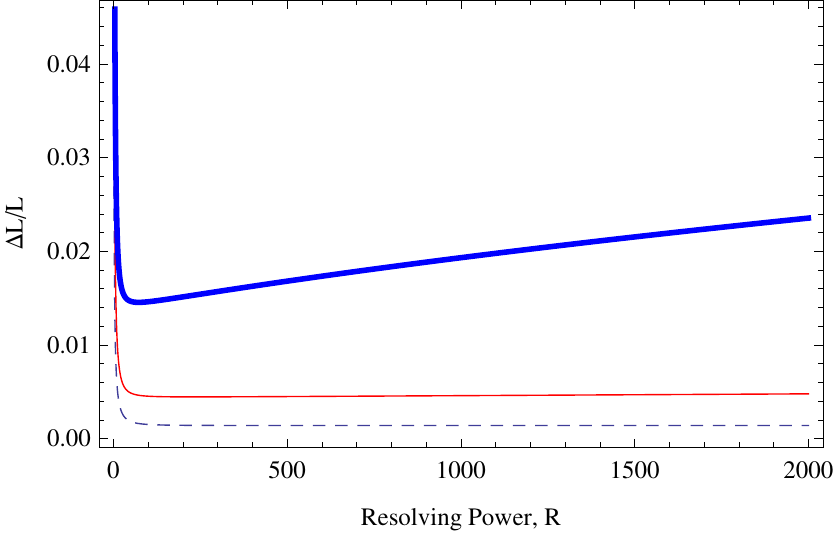}
\caption{RPE in luminosity from turbulent dissipation  for optically thin case at $r=10$AU for the parameters
of LLRL 31 as a function of resolving power $R_{ef}$ for $\alpha = 0.1 (\text{Thick lined Blue}), 0.01 (\text{thin lined Red}), 0.001 (\text{Dashed})$. The higher $\alpha$ has the most prominent minimum, which does not correspond to the maximum resolving power but an intermediate value.}
\label{resolvingpower}
\end{figure}

%How should we design our telescope, to minimize the resolution problem?
%We must combine Equations~\ref{eq:xioptdefine} and~\ref{eq:xirealdefine}, ensuring that $R$ is chosen to make $\xi_{\text{opt}} = \xi_{\text{real}}$ at all radii.
%This is non-trivial, since it is obvious from Equation~\ref{eq:xioptdefine} that the optimal value for $R$ varies with the structure of the disc - the very thing we really wish to measure!
%It should also be remembered that we have made the somewhat extreme assumption that each annulus of material within the disc has a well characterized temperature, and that emission from that annulus occurs at a single frequency.

%The behavior of Equation~\ref{eq:LvaryReal} as the telescope resolution, $R$, increases highlights the subtleties of accretion disc theory; subtleties which are usually swept under a small rug emblazoned with $\alpha$.
%Increased resolution is normally thought of as a Good Thing, but this is not necessarily the case when one is attempting to extract a mean field theory from a turbulent disc.
%Too much resolution will cause data points to be taken from individual eddies, rather than averaged over a number of eddies.
%It should also be remembered that we have assumed that the disc is in a temporarily steady state - we do not permit the accretion rate to vary with radius.
%This is unlikely to be the case for a real disc, which would further complicate our analysis.
%
%
%-------------------------------------------
%
%
\section{Implications for Multi-epoch Observations}

\begin{figure}
\includegraphics{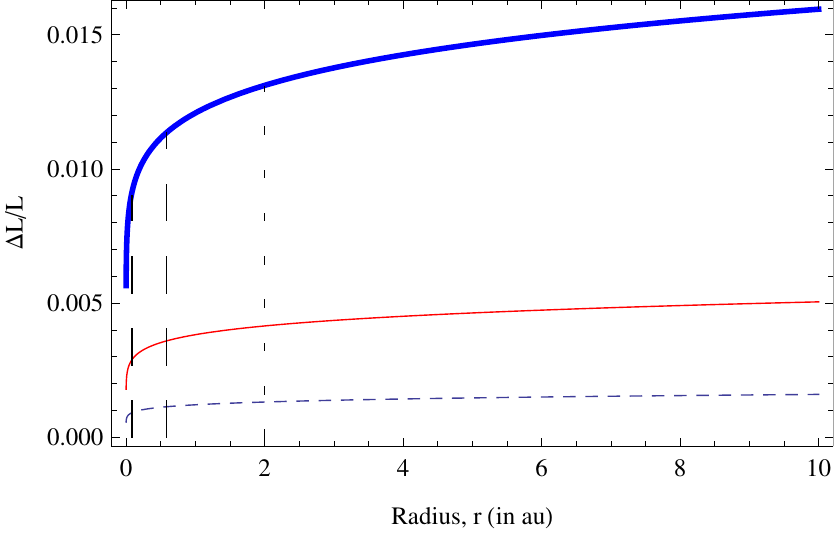}
\caption{RPE in luminosity from turbulent dissipation vs. radius $r$ for the optically thin case, using  $\alpha = 0.1 (\text{thick lined Blue}), 0.01 (\text{thin lined Red}), 0.001 (\text{Dashed})$. From left to right the three vertical dashed lines
correspond  to three specific radii whose orbit period equals different inter-epoch observation periods for  LRLL 31
\citep{2009ApJ...704L..15M}: 1 week ($r = 0.087$ AU) and 4 months($r = 0.58$ AU), and
 $\sim$ 25 months ($r=2$ AU).  For the respective inter-epoch periods,  these vertical lines mark the upper radial limit of observational validity of the curves, and above which the expected variability will be sharply reduced.
For example, if the inter-epoch interval is 25 months, then the RPE would predict the regime of validity for the curves to be $r< 2\AU$. Above $r> 2$ AU the prediction would be that $\Delta L_\nu/L_\nu$ falls near zero because the epochs would sample data from the same statistical ensemble.}
\label{fig:variabilityvsr}
\end{figure}

A basic application of the RPE  described above arises for multi-epoch observations
of a single object when the time scale for a  snapshot observation is shorter than
the eddy time at any radius. Emission from  radii
at which the time scale between epochs of observation
 exceeds the eddy turnover time  would then be expected to vary stochastically between epochs.
In contrast,  at radii larger than the that at which the eddy turnover time matches the epoch
interval time, successive observations sample the same member of the statistical
ensemble and no stochastic variability corresponding to the the local energy-dominating eddies would be expected.
%EGB: turbulence is actually a spectrum. we may want to generalize in future

\subsection{Lessons from the specific case  of LRLL 31}

Presently, multi-epoch data are sparse. Programs such as YSOVAR \citep{2010AAS...21542907S} will
eventually produce more data.  Nevertheless, to focus the practical application on a
on a specific object, we consider  IR Spitzer multi epoch observations
of the source LRLL 31 in the star-forming region IC 348
\citep{2009ApJ...704L..15M}
.  For reasons discussed below, this object
is most likely not showing its primary variability from  stochastic effects  we discuss but our analysis
to reach this conclusion below  illustrates how to use the concepts of the present paper to arrive at that conclusion.

 The stellar mass of LRLL 31 is $M = 1.8 M_{\odot}$ and the estimated accretion rate is
 $\dot{M} = 1.5 \times 10^{-8} M_{\odot} {\rm  yr}^{-1} ({\rm from\ }\text{Pa} \beta)$.
 %This object is a transitional disc with a dust depleted, optically thin inner region.
%There is also data for some object called Br \gamma but there is only a very small difference.
 For the low spectral resolution mode used to observe this source
 in the 5-38 micron range \citep{2009ApJ...704L..15M},
 the Spitzer spectral resolving power is $R\sim 200$. The observation times are minutes, well below
 the smallest relevant orbit time and consistent with setting $t_{obs}\sim 0$ in
 (\ref{12a}) and (\ref{eq:NflDefine}) and out assumption throughout the previous sections.

%EGB lets check, maybe 100

Figure 2 shows Eq. (\ref{eq:LvaryReal}) plotted as a function of  radius for three different values of
$\alpha$ and a value of $R_{ef}= 200$ for the optically thin case.
%EGB was R=200 used here?
 %EGB what choice of p is used?
  The vertical lines show, for three separate choices of the inter-epoch time,
   the radii above which there would be no variability expected because the inter-epoch time is less than the eddy turnover time above these radii.

Tables 2  and  3  exemplify  how use of $R_{ef}=R$,
the resolving power for the telescope, compares to
the use of $R_{ef}=R_{opt}$, the value that  minimizes the luminosity RPE,
% in calculating
%(\ref{eq:LvaryRealCom})
 for different choices of $r$ and $\alpha$.
For these tables, the epoch time is taken to exceed
the eddy turnover time at all radii. In reality, as emphasized by the vertical lines in Figure 2,
there would be negligible expected variability at radii above which the orbit time ($\sim$ eddy turnover time) exceeds a chosen inter-epoch observation interval.
For cases  in which  the telescope resolution (column 6) exceeds the optimal value (column 4)
the data can  in principle be binned such that  the binned data can be compared to the optimal $\alpha$-disc model prediction.  The tables highlight two trends in the Tables:  (i) the larger the value of $\alpha$ the larger the predicted
variability between epochs at a given radius and (ii) the larger the radius for a fixed $\alpha$ the larger the predicted variability.
%and (iii) the larger the radius the larger the difference between the variability at $R_{opt}$ and
%$R=200$, where the is determined by the telescope.

Muzerolle (2009) argue  that the disc of LLRL 31,  at  $ r < 10$ AU,
 is optically thin and substantially cleared out (perhaps via the action of planet) below $10$AU but
optically thick for a least some range of larger radii. Thus for this object,
the values in the Table 2 would apply for $r<10$AU and the values of table 3 may apply for
some range of larger radii above which the disc  again becomes optically thin.
The value at which the latter occurs requires
more detailed modeling.
% and has to be modeled and reconciled with that predicted
%in theoretical models \citep{2007MNRAS.381.1280E}.

% the values for $r = 0.1 {\rm AU} - 10 {\rm AU}$, we have used $p = 0.6$ and the relevant $h/r$ from Eq. 32.
%For $r = 20$ AU, the disc is presumabl optically thick and so $p = 0.9$ and the relevant equation from Eq. 32 is used.
%Complementarily,
 %Eqn. (8) of \cite{2007MNRAS.381.1280E}
%identifies a  transition  radius
%\beq
%r_{th} = 1.2 AU \left(\frac{\dot{M}_{\ast}}{10^{-8} M_{\odot} yr^{-1}} \right)^{\frac{2}{3}} \left(\frac{M_{\ast}}{M_{\odot}} \right)^{\frac{1}{3}} \left(\frac{\alpha}{0.01} \right)^{-\frac{2}{3}}
%\eeq
% below which an otherwise isolated protplanetary disc would be  optically thick and above which it becomes  thin.
%The combination of this along with the inner hole imply that for all radii, Table 2, the optically thin
%regime is the more appropriate for computing the RPEs.

 There is also  a transition radius
above which viscous dissipation would be dominated by irradiation from the central star \citep{2006ApJ...638..897S,2007MNRAS.381.1280E},
% namely \cite{2007MNRAS381.1280E}
namely
\beq
r_{c} = 0.3 AU \left(\frac{\dot{M}_{\ast}}{10^{-8} M_{\odot} yr^{-1}} \right)^{\frac{3}{4}} \left(\frac{M_{\ast}}{M_{\odot}} \right)^{\frac{7}{6}} \left(\frac{L_{\ast}}{L_{\odot}} \right)^{-\frac{5}{6}}.
\eeq
%For $1.8M_{\odot}$, $\alpha = 0.01$, $L_* = 3 L_{\odot}$ and $\dot(M) = 1.5 \times 10^{-8}M_\odot$/yr, this viscous to irradiation transition occurs at $R_{tr} = 0.32$ AU whereas for $L_*=60L_\odot$, $r_{tr}=0.0266$ AU).
%Note that  the best fit protoplanetary disc models suggest that only inside some critical
%radius $r_c$ (whose value is  $r_c< 1$AU given the intermediate
%accretion rate of LRLL 31 and mass larger than the 0.5$M_\odot$ fiducial case of D'Alessio et al. 1999)
For $r>r_c$  the photospheric temperature induced by viscous dissipation falls
below that associated from surface dust illumination by the central star.
The values of $p$ would then flatten to $p=1/3$ (see Table 1) and irradiation would dominate
at all larger radii.
%\cite{2006ApJ...638..897S}.
Disc emission resulting from  stellar irradiation
 is less influenced by the stochasticity of turbulence than emission via turbulent dissipation.
Since the local blackbody emission varies with $T^3$, we would expect a much reduced stochastic variability in regions of the disc dominated by emission from reprocessed starlight.
This implies that even for  epoch
intervals longer than the orbit periods at $r>r_c$ that the stochastic variability
discussed herein would be reduced compared to regions of $r<r_c$
It is important to emphasize that the RPEs calculated in  Tables 2 and 3  is a stochastic variability
corresponding to that of  mean field $\alpha$-disc
resulting from the disc luminosity contribution by stochastic process
such as  turbulent viscous dissipation. As applied to the specific object LLRL 31 therefore,
these would only amount to a variability in a subdominant contribution to the total emission above $r_c$.

 An absence of observed stochastic fluctuations
 for $r>r_c$ would support  the expectation that the primary emission in those regions is not the result of turbulent dissipation. For $r<r_c$ the absence of stochastic variability would be expected only
if  the inter-epoch time scale is shorter than the characteristic eddy turnover time at the radius
dominating the radiation at the particular frequency measured.
Note also that  the amplitude of the predicted stochastic variability
would  not depend on the epoch interval as long as the epoch interval is larger than the orbit time
at the radius producing the corresponding emission.
Other systematic non-axisymmetry or variable accretion rates would  add to the variability and could
be distinguished form stochastic effects given enough multi-epoch observations

For the specific case of LRLL 31, the largest time scale between epochs  is of order 4 months
which means that according to Fig. 2, the stochastic variability would be expected
only for $r< 0.58$ AU. This is well within the dust depleted inner region of this transitional object where
the disc is optically thin and where  the emission would come from dissipation internal to the disc rather than
reprocessed starlight.  Thus there would be some stochastic contribution to the measured variability between epochs predicted by the RPE of the mean field models.
% The observed variability in this source is larger for lower wavelengths which does correspond to the trend that  we would predict because the longer wavelengths are produced at larger radii, above the critical radius where the epoch time is less than the orbit period.
{ However,  the double-digit percentages of  the observed variability reported by
\citet{2009ApJ...704L..15M}
 in LRLL 31 are larger than the values predicted in our Tables that arise  from stochastic variability alone }
 %This is  exacerbated by the fact that for discs with inner holes, the emission would also include contributions from eddies dissipating below the top layer of eddies, unlike for an optically thick disc. Retaining the   optical depth in  $N_{ed}$, $N_{rs}$, and $N_{fl}$ using the first equations of
%Eqs. (\ref{2}), (\ref{12a}), (\ref{eq:NflDefine}) leads to inverse powers of $\alpha$ appearing in
%(\ref{30thin})
% which somewhat increase the expected variability for the optically thin case compared to the optically thick
%case of (\ref{eq:NflDefine}).
In addition, the presence  of a single pivot wavelength  of 8.5 microns (see Fig 1 of
\citet{2009ApJ...704L..15M})
  at which there is no variability in this source
and above which the sign of the variability between epochs changes sign from that at low wavelengths
strongly suggests the influence of a global geometric feature such as a thick wall or a warp.
 In short
the dominant variability in LRLL 31 object likely comes from systematic rather than stochastic effects.

%r_{2} = 0.3 AU \left(\frac{\dot{M}_{\ast}}{10^{-8} M_{\odot} yr^{-1}} \right)^{\frac{3}{4}} \left(\frac{M_{\ast}}{M_{\odot}} \right)^{\frac{7}{6}} \left(\frac{L_{\ast}}{L_{\odot}} \right)^{-\frac{5}{6}}\eeq
%For $1.8M_{\odot}$, $\alpha = 0.01$, $L_* = 3 L_{\odot}$ and $\dot(M) = 1.5 \times 10^{-8}M_\odot$/yr, this viscous to irradiation transition occurs at $R_{tr} = 0.32$ AU whereas for $L_*=60L_\odot$, $r_{tr}=0.0266$ AU).

%\citep{2009ApJ...704L..15M} shows,

%For $\Delta t_{\text{epoch}} = 4 \text{months}$,
% the variability  is 30\% at long wavelengths, 60\% at short and
%$\Delta t_{\text{epoch}} = 1 \text{week}$ difference is 20-30\%.

%EGB: I think Resolving power R= ~100 for spitzer in the "low" mode used by Muzerolle for IRS 5-38 micron.--need reference.
%for 0.8-2.5 microns looks like 1200 spectral resolution. see section 2.2

%\subsection{Summary of Implications for Future IR Multi-Epoch Disc Observations}

%The main point of our work here is to  quantitative the amount of intrinsic stochastic variability
%expected when using a mean field accretion disc model to compare with data.
%This is meant to be a tool for helping to distinguish sources of stochastic variability from systematic
%variability.  Presently
%In this respect a summary of the predictions are

\section{Conclusions}

For $\alpha$  models of turbulent accretion discs we have derived a predicted
RPE in the luminosity as  function of frequency, focusing on the case
in which the observational exposure times  are small compared ti other dynamical time scales.
We  have discussed the implications for interpreting multi-epoch spectral observations of spatially unresolved protoplanetary discs.

The RPE depends on the radial scale of averaging and there exists an
optimal scale that minimizes the error which also corresponds to an optimal
spectral resolution when the correspondence between peak frequency of emission and radius
is made.  The data can be binned to compare with the theory of minimum
RPE  only when the spectral resolution of the telescope exceeds the value $R_{opt}$
 which minimizes the RPE.
 An optimal resolution exists because if the spectral resolution is too high then the instrument ends up
 sampling noise and if the spectral resolution is too low, then the instrument
 samples overly coarsely binned regions of the disk.

 The stochastic variability expected between multi-epoch observations of discs is an implicit prediction of
alpha disc theory and  can be directly compared to observed variabilities.
Complementarily,  the nature of the observed variability can  be used to constrain whether or not the behavior
of the disc is consistent with luminosity produced by stochastic or systematic processes.

The predicted stochastic  variability from $\alpha$ discs has several distinct characteristics that would signature its prevalence:

$\bullet$   The stochastic variability is predicted primarily for regions where the luminosity is dominated by internal
turbulent dissipation in the disc. For regions dominated by surface dust-reprocessed starlight, the stochastic variability could be expected to be strongly reduced.

$\bullet$.  For regions of the discs dominated  by from turbulent dissipation,
the strength of the stochastic variabilty would increase gradually with decreasing emission
frequency (and thus increasing radius) down to a critical frequency
below which the associated disc radii have orbital times exceeding the inter-epoch observation time. There the expected variability from the RPE would drop sharply.

$\bullet$  For inner regions of the discs in transitional objects, where emission may be dominated by
turbulent dissipation but the discs are optically thin, the stochastic variability would be expected
to be lower than that expected from the same radii for optically thick turbulent dissipation dominated discs.

In applying the basic ideas herein to the specific case of LRLL 31, we find that systematic variability NOT stochastic variability dominates.  This is indeed consistent with the conclusion of \citet{2009ApJ...704L..15M}.
As more multi-epoch observations of broader samples of discs at different stages in their lifetimes
are obtained,  the present work  may help provide a tool to distinguish stochastic processes
in discs from systematic dynamical changes and the time scales on which these occur.

Two overall lessons from this analysis are: (1) the precision of axisymmetric accretion theories can be quantified and deviations the alpha-accretion disc theory
and observations at a given epoch cannot rule out the theory if the deviations between
epochs exhibit stochastic behavior and fall within the expected RPE of the theory.
The theory may be incomplete but because a mean field theory for a turbulent system is intrinsically imprecise,
but the imprecision must be quantified so that the user realizes its limitations. (2)
Arbitrarily high spectral resolution can lead to  misleading comparisons between
theory and observation and binning the data may be necessary to compare the data with the theory of
minimum RPE.

\begin{table}
\begin{tabular}{c|c|c|c|c|c}
r(au) & $\alpha$ & $R_{\text{opt}}$ & ${\Delta L_\nu \over L_\nu}$
(\%) & $R_{\text{tel}}$ & ${\Delta L_\nu \over L_\nu}$(\%) \\
\hline
0.1 & 0.001 & 1832 & 0.08754 & 200 & 0.09226 \\
& 0.01 & 579.3 & 0.2781 & 200 & 0.2816 \\
& 0.1 & 183.2 & 0.8914 & 200 & 0.8915 \\
1 & 0.001 & 1156 & 0.1103 & 200 & 0.1139 \\
& 0.01 & 365.5 & 0.3514 & 200 & 0.3531 \\
& 0.1 & 115.6 & 1.135 & 200 & 1.144 \\
10 & 0.001 & 729.3 & 0.1392 & 200 & 0.1417 \\
& 0.01 & 230.6 & 0.4449 & 200 & 0.4451 \\
& 0.1 & 72.93 & 1.454 & 200 & 1.513 \\
20 & 0.001 & 634.9 & 0.1493 & 200 & 0.1515 \\
& 0.01 & 200.8 & 0.4780 & 200 & 0.4780 \\
& 0.1 & 63.49 & 1.570 & 200 & 1.661 \\
\end{tabular}
\caption{Optically Thin Case.
In column 4, we list the RPE varibilities calculated using $R_{ef}=R_{opt}$ and
and in column 6  the values corresponding to the Spitzer telescope low mode   $R_{ef}=R_{tel}\sim  200$.
The RPEs  are computed from Eq. (\ref{30thin}) for different $r$ and $\alpha$ and  $h/r$ from Eq. (\ref{depththin}) using $p=0.6$.
The optimal resolution is calculated from
Eq. (\ref{eq:xioptdefine})  as described below
Eq. (\ref{depththin}).
%from Equations (\ref{eq:xirealdefine}) and (\ref{eq:xioptdefine})
%EGB Here again need to check that value of R_opt from minimizing 27 really corresponds to same value  in 22. the minimization could occur at two different values.
%The actual resolution \citep{2009ApJ...704L..15M} ranges from 200-400 but we use the lowest value to look at the minimal variability.
Note that if $R_{opt}< R_{tel}$ data can be binned to compare with the $\alpha$-disc model of minimal RPE.}
\label{tbl:variabilityalpha}
\end{table}
\begin{table}
\begin{tabular}{c|c|c|c|c|c}
r(au) & $\alpha$ & $R_{\text{opt}}$ & ${\Delta L_\nu \over L_\nu}$(\%) & $R_{\text{tel}}$ & ${\Delta L_\nu \over L_\nu}$(\%) \\
\hline
0.1 & 0.001 & 625.8 & 0.6935 & 200 & 0.7111 \\
& 0.01 & 249.2 & 1.116 & 200 & 1.118 \\
& 0.1 & 99.19 & 1.835 & 200 & 1.881 \\
1 & 0.001 & 557.8 & 0.7355 & 200 & 0.7510 \\
& 0.01 & 222.1 & 1.186 & 200 & 1.186 \\
& 0.1 & 88.40 & 1.958 & 200 & 2.030 \\
10 & 0.001 & 497.1 & 0.7802 & 200 & 0.7935 \\
& 0.01 & 197.9 & 1.260 & 200 & 1.260 \\
& 0.1 & 78.79 & 2.091 & 200 & 2.200 \\
20 & 0.001 & 480.2 & 0.7942 & 200 & 0.8069 \\
& 0.01 & 191.2 & 1.284 & 200 & 1.284 \\
& 0.1 & 76.10 & 2.133 & 200 & 2.256 \\
\end{tabular}
\caption{same as Table 2 but for optically thick case. The RPEs  are computed from Eqs. (\ref{eq:LvaryOptCom}) and  (\ref{eq:LvaryRealCom}) for different $r$ and $\alpha$ and for the optically thick version of $h/r$ in Eq. (\ref{depth}) and $p=0.9$.}
\label{tbl:variabilityalphathick}
\end{table}
% ------------
%
% Bibliography

\bibliography{general}

\begin{thebibliography}{}

\bibitem[\protect\astroncite{{Balbus} and {Hawley}}{2003}]{2003LNP...614..329B}
{Balbus}, S.~A. and {Hawley}, J.~F.: 2003,
\newblock in {E.~Falgarone \& T.~Passot} (ed.), {\em Turbulence and Magnetic
  Fields in Astrophysics}, Vol. 614 of {\em Lecture Notes in Physics, Berlin
  Springer Verlag}, pp 329--348

\bibitem[\protect\astroncite{{Blackman}}{1998}]{1998MNRAS.299L..48B}
{Blackman}, E.~G.: 1998,
\newblock {\em \mnras} {\bf 299}, L48

\bibitem[\protect\astroncite{{Blackman}}{2010}]{2010AN....331..101B}
{Blackman}, E.~G.: 2010,
\newblock {\em Astronomische Nachrichten} {\bf 331}, 101

\bibitem[\protect\astroncite{{Edgar} et~al.}{2007}]{2007MNRAS.381.1280E}
{Edgar}, R.~G., {Quillen}, A.~C., and {Park}, J.: 2007,
\newblock {\em \mnras} {\bf 381}, 1280

\bibitem[\protect\astroncite{{Frank} et~al.}{2002}]{2002apa..book.....F}
{Frank}, J., {King}, A., and {Raine}, D.~J.: 2002,
\newblock {\em {Accretion Power in Astrophysics: Third Edition}},
\newblock Cambridge University Press

\bibitem[\protect\astroncite{{Hartmann}}{2009}]{2009apsf.book.....H}
{Hartmann}, L.: 2009,
\newblock {\em {Accretion Processes in Star Formation: Second Edition}},
\newblock Cambridge University Press

\bibitem[\protect\astroncite{{Muzerolle} et~al.}{2009}]{2009ApJ...704L..15M}
{Muzerolle}, J., {Flaherty}, K., {Balog}, Z., {Furlan}, E., {Smith}, P.~S.,
  {Allen}, L., {Calvet}, N., {D'Alessio}, P., {Megeath}, S.~T., {Muench}, A.,
  {Rieke}, G.~H., and {Sherry}, W.~H.: 2009,
\newblock {\em \apjl} {\bf 704}, L15

\bibitem[\protect\astroncite{{Pringle}}{1981}]{1981ARA&A..19..137P}
{Pringle}, J.~E.: 1981,
\newblock {\em \araa} {\bf 19}, 137

\bibitem[\protect\astroncite{{Shakura} and
  {Sunyaev}}{1973}]{1973A&A....24..337S}
{Shakura}, N.~I. and {Sunyaev}, R.~A.: 1973,
\newblock {\em \aap} {\bf 24}, 337

\bibitem[\protect\astroncite{{Sicilia-Aguilar}
  et~al.}{2006}]{2006ApJ...638..897S}
{Sicilia-Aguilar}, A., {Hartmann}, L., {Calvet}, N., {Megeath}, S.~T.,
  {Muzerolle}, J., {Allen}, L., {D'Alessio}, P., {Mer{\'{\i}}n}, B.,
  {Stauffer}, J., {Young}, E., and {Lada}, C.: 2006,
\newblock {\em \apj} {\bf 638}, 897

\bibitem[\protect\astroncite{{Spruit}}{2010}]{2010arXiv1005.5279S}
{Spruit}, H.~C.: 2010,
\newblock {\em ArXiv e-prints}

\bibitem[\protect\astroncite{{Stauffer} et~al.}{2010}]{2010AAS...21542907S}
{Stauffer}, J.~R., {Megeath}, T., {Rebull}, L., {Morales}, M., {Plavchan}, P.,
  {Gutermuth}, R., {Inseok}, S., {Carey}, S., {Covey}, K., and {YSOVAR/Orion
  team}: 2010,
\newblock in {\em Bulletin of the American Astronomical Society}, Vol.~41 of
  {\em Bulletin of the American Astronomical Society}, pp 350--+

\end{thebibliography}
\bibliographystyle{astron}

% ------------

\section*{Acknowledgments}

The authors acknowledge support from NSF grants AST-0406799, AST-0098442, AST-0406823, and NASA grants ATP04-0000-0016 and NNG04GM12G (issued through the Origins of Solar Systems Program). FN acknowledges
a Horton Fellowship from the Laboratory for Laser Energetics at U. Rochester.

\bsp

\label{lastpage}

\end{document}